\documentclass[journal]{IEEEtran}
\usepackage{amsmath,amsfonts}
\usepackage{algorithmic}
\usepackage{algorithm}
\usepackage{array}
\usepackage[caption=false,font=normalsize,labelfont=sf,textfont=sf]{subfig}
\usepackage{textcomp}
\usepackage{stfloats}
\usepackage{url}
\usepackage{verbatim}
\usepackage{graphicx}
\usepackage{cite}
\usepackage{makecell}
\usepackage{amsmath}

\usepackage{amsthm}

\usepackage{hhline} 
\usepackage{multirow}
\usepackage[table]{xcolor}

\usepackage[shortlabels]{enumitem} 

\usepackage{amsfonts}
\newcommand{\reals}{\Re}

\newcommand \inpSet {U}
\newcommand \inpPt {u}
\newcommand \inpSig {\vec{\inpPt}}
\newcommand{\stPt}{x}
\newcommand \sttraj {\vec{\stPt}}
\newcommand \formula \phi 

\renewcommand{\Re}{\mathbb{R}}
\theoremstyle{plain} 
\newtheorem{theorem}{Theorem}[section]

\theoremstyle{definition}
\newtheorem{definition}{Definition}[section]

\begin{document}

\title{Privacy-Preserving Robustness-Guided Testing%
\thanks{All authors affiliated with Oregon State University at time of submission. Emails: \{kollch,houssam.abbas,mike.rosulek\}@oregonstate.edu. \\ Research partially supported by NSF award 2118179.}%
}

\author{Charles Koll, Houssam Abbas, and Mike Rosulek}

\markboth{Journal of \LaTeX\ Class Files,~Vol.~14, No.~8, August~2021}%
{Shell \MakeLowercase{\textit{et al.}}: A Sample Article Using IEEEtran.cls for IEEE Journals}


\maketitle

\begin{abstract}
In industries like semi-conductors and aerospace, it is common for the design company to contract out part of the testing process to a Testing-as-a-Service (TaaS) company. In such a setup, both parties want to keep their intellectual property private (i.e. unknown to the other party): namely the design and its output traces for the designer, and the detailed correctness specifications for the tester.
We demonstrate a practical way for conducting provably private TaaS using a cryptographically secure multi-party computation protocol. We do so for the common setup where the specification is in Signal Temporal Logic, and testing is done using robustness-guided falsification. Our implementation carefully separates the steps of falsification, and manages to be \textit{maximally secure}, in the sense that the parties learn the absolutely minimum necessary for falsification.
For up to a 100-dimensional input search space, the generation of one test takes less than 80 msec, and the overhead of the privacy-preserving protocol is less than 6x. These numbers mean that privacy-preserving TaaS is practical, thus enabling TaaS in security-sensitive industries.
\end{abstract}

\begin{IEEEkeywords}
Multi-Party Computation, falsification, cyber-physical systems, privacy-preserving computation.
\end{IEEEkeywords}

\section{Introduction}
{\em Testing} remains the most widely used technique to ensure embedded and cyber-physical system correctness.
\emph{Testing-as-a-Service} (TaaS), which is the practice of outsourcing some testing activities to an outside party, is common in industries like semi-conductor, aerospace, fuel cell batteries, and software.
In such a setup, the testing company, aka Tester, possesses detailed correctness specifications which it developed over years of experience testing different designs, e.g. many kinds of plane engines. This constitutes Tester's intellectual property and so Tester wishes to keep it secret.
The design company, aka Designer, is of course incentivized to keep the System Under Test (SUT) secret. In particular, this implies keeping the SUT's output traces secret, since otherwise Tester might learn to map input traces (which it generates) to output traces. 
\textit{A key challenge of TaaS is, therefore, the protection of the intellectual properties (IPs) of both the design and the testing companies}.

Currently, Designer and Tester sign Non-Disclosure Agreements (NDAs), which provide legal penalties if the parties leak each other's IP to someone else. 
However, a much stronger guarantee would be if 1) each party's IP is never shared with the other party to begin with, and 2) this guarantee is \textit{technological}, rather than legal. I.e. we want to make it \textit{impossible} for the parties to learn each other's IP, \textit{while still enabling them to jointly perform the testing of the SUT}. We call this \textit{privacy-preserving testing}.

This paper presents a practical way to perform private testing for cyber-physical systems (CPS) in the following common setting.
The specification is expressed in Signal Temporal Logic (STL) \cite{maler2004monitoring} (or similar logic); STL is a dense-time logic allowing the specification of real-time constraints, like `The speed increases within 2.1sec of the distance exceeding the threshold'.
Given an output trace $x$ from the SUT, the degree to which it satisfies or violates the STL specification $\formula$ is formalized using the notion of \textit{robustness} $\rho(x,\formula)$ \cite{fainekos2009robustness}: if positive, this indicates $x$ satisfies $\formula$, and if negative, this indicates that $x$ violates $\formula$, while its magnitude $|\rho(x,\formula)|$ indicates how well it satisfies, or how badly it violates, $\formula$. 
The test generator's task, then, is to search the SUT's input space for inputs $u$ s.t. the corresponding output trace $x$ has negative robustness. This search is done using optimization heuristics, most commonly simulated annealing \cite{kirkpatrick1983optimization} or some extension \cite{xiang1997generalized}. The robustness values guide the test generator towards promising inputs.
This entire process is called \emph{falsification}.

We develop a privacy-preserving simulated annealing (SA)-based implementation of falsification using a Multi-Party Computation (MPC) \cite{evans2018pragmatic}, which is a class of cryptographic protocols with rigorous privacy guarantees. Privacy means that the probability that Designer will learn \textit{anything} about Tester's formula can be made arbitrarily small (in the security parameter), and similarly for the probability of Tester learning \textit{anything} about the SUT's traces. (Security definitions are given in the next section).
In fact, our implementation \emph{hides all that can be hidden}: the Tester does not even learn the inputs that it is generating to simulate the SUT, and neither party learns the robustness values that are guiding the falsification. Designer learns the input signals since it needs to apply them to the SUT.
The reason for this maximal privacy is that it protects the TaaS against future adversaries that might figure out a way to leverage, say, knowledge of the robustness values. Since these values remain cryptographically protected, no adversary can use them.
This maximal privacy goes beyond the private computation of robustness values: every iteration of an SA loop requires a comparison of robustness values to select a new input signal, so hiding this comparison while allowing Designer to learn the input signals is non-trivial.

We show experimentally that our private falsification is practical and does not incur exorbitant overhead relative to in-the-clear (non-private) falsification.

\textbf{Related work.}
There is no evident prior work on privacy-preserving testing. However, private monitoring, a component of private testing, has a few works.
Abbas \cite{abbas19prv} introduced the concept of private offline monitoring for LTL, and suggested several approaches to preserving the privacy of the monitored signal, but not of the LTL formula.
We work with Signal Temporal Logic (STL), interpreted over dense-time signals, necessary for cyber-physical systems. STL does not have an equivalent finite automaton representation, so works like \cite{sasakawa2014oblivious} and \cite{waga2025oblivious}, which work in discrete time and use finite automata, are not applicable.
Henzinger et al. \cite{henzinger2025privacy} approached private runtime verification (rather than testing) by focusing on simple functionality such as adders and comparators, and encoded the specification as a circuit -- our testing functionality takes the specification as a private input.
Karimi et al. \cite{karimi2026sharing} privately monitor over a distributed monitor, where at least one distributed component is honest (does not attempt to learn private information) -- our work assumes that both parties attempt to learn all that they can, while following the protocol.

Some works implement SA as a multi-party computation. The most similar work is \cite{sakuma2026scalability}, which implements SA in MP-SPDZ for $n$ parties, but requires three parties for its protocol and uses a discrete space for candidate points. 
Our implementation operates over continuous space, namely, the space of possible CPS control inputs.
\cite{hong2014collaboratively} also operates on a discrete space, and further does not hide SA's intermediate function comparisons, leaking statistical information regarding private data. 
Our implementation must hide intermediate comparisons.
\cite{silaghi2005secure} likewise reveals intermediate function comparisons.
Unlike these works, our private SA operates over a continuous space and does not reveal the SA function comparisons, achieving a maximally private privacy-preserving testing.

\textbf{Contributions.}
Our contributions are
1) an MPC for robustness-guided falsification that keeps private the maximum possible; and
2) an experimental evaluation of our implementation, demonstrating that secure multi-party falsification remains practical despite the overhead of privacy-preserving computations.

\section{What is Multi-Party Computation?}
Secure Multi-Party Computation (MPC) is a branch of cryptography that enables multiple parties to together compute some functionality without revealing their private inputs. We consider two parties, Alice and Bob, holding private data $x$ and $y$ respectively. Together they want to compute some functionality $f(x,y)$, but Alice does not want Bob to learn $x$, and Bob does not want Alice to learn $y$. An example of this could be computing robustness $f(x,y)$ of a dense-time signal $x$ with respect to an STL formula $y$.
Using a \emph{secure} or \emph{privacy-preserving interactive MPC protocol}, they can follow a sequence of steps to compute and learn the result of $f(x,y)$ \emph{(Correctness)} and learn nothing else \emph{(Privacy)}. Rigorous definitions of ``learning nothing else'', input sizes, models of parties, etc. have been thoroughly addressed in the MPC literature (e.g. see the book \cite{Cramer_Damgrd_Nielsen_2015}). A set of self-contained security definitions is beyond the scope of this paper, but we provide the following informal statement:
\begin{definition}[Secure protocol (Intuitive)]
\label{def:intuitive security}
    An MPC protocol implementing functionality $f$ is considered \emph{secure} w.r.t. semi-honest parties (i.e., parties that follow the protocol but try to learn extra information based on the messages they see) if it is highly improbable that the parties could learn anything extra from executing the protocol that they couldn't learn anyway, just from knowing their inputs and the final value $f(x,y)$.
\end{definition}

The following is a well-known MPC result.
\begin{theorem}
\label{thm:general feasibility}
    Any functionality $f$ that can be computed in time that is polynomial in the input sizes can also be computed by any secure interactive MPC protocol.
\end{theorem}

This theorem establishes that theoretically, secure MPC can provide the privacy guarantees necessary for privacy-preserving robustness-guided falsification.
There are indeed general-purpose \emph{compilers} that can implement such an interactive protocol from a description of $f$.
In this work, we use one such compiler called MP-SPDZ \cite{mp-spdz}, which can accept $f$ specified in a subset of Python.
While compilers like MP-SPDZ provide a ready-made solution for MPC implementation, the resulting protocol is often prohibitively inefficient in practice.
It is commonly necessary to tailor the computation $f$ to be more ``MPC-friendly.''
In this work we do this by identifying isolated components of computation that can be handled separately of each other without breaking privacy guarantees. We elaborate this in the following section.

\section{Private Robustness-Guided Falsification}
\begin{figure}
    \centering
    \includegraphics[width=0.9\linewidth]{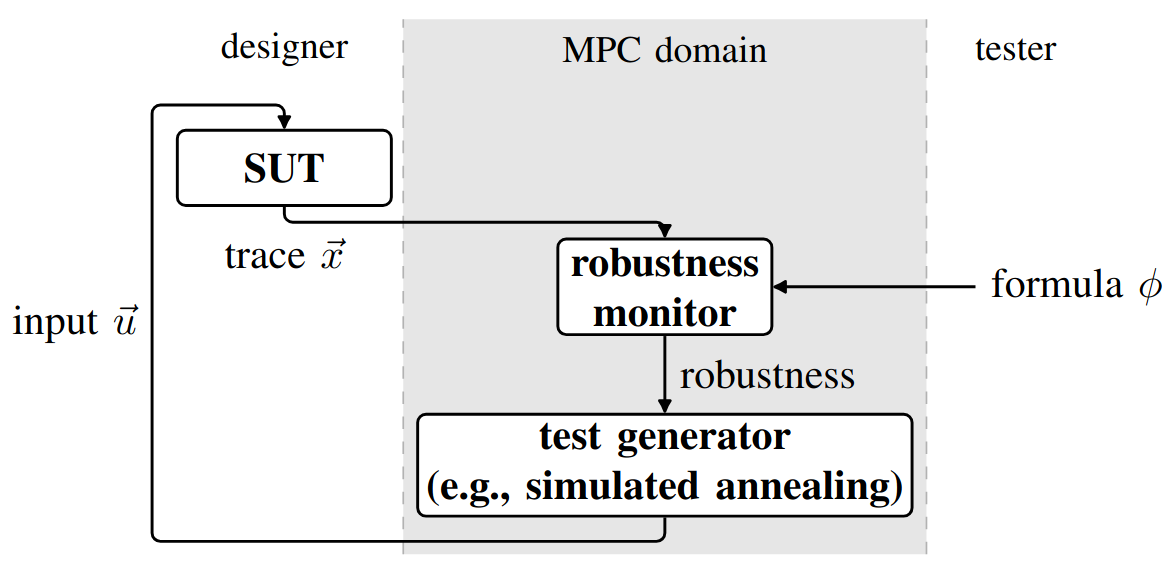}
    \caption{Private falsification. Designer contributes a (private) system trace $\sttraj$ and Tester contributes a (private) formula $\formula$. Robustness computation, and the generation of the next test based on $\rho(\sttraj,\formula)$, happen in MPC. The next test is revealed to Designer only, who inputs it to the SUT, and the loop repeats.}
    \label{fig:MPC testing overview}
\end{figure}

\textbf{Testing loop overview.}
As explained in the introduction, falsification proceeds by minimizing robustness over the space of valid input signals to the SUT. A negative minimum robustness indicates a bug that should be debugged. 
If we let $\inpSig \in \inpSet^m \subset \reals^m$ denote the discretized $m$-dimensional input signal that is fed to the SUT and $S$ denote the SUT which maps input $\inpSig$ to output trace $\sttraj$, then falsification solves
\begin{equation}
    \label{eq:falsification}
    \min_{\inpSig \in \inpSet^m} \rho(S(\inpSig),\formula)
\end{equation}
Since $S$ is typically a black box (we can only simulate it but not formally analyze it), this minimization is solved using blackbox optimizers. We focus on one of the optimizers most commonly used in falsification, Simulated Annealing (SA)~\cite{kirkpatrick1983optimization}.
As shown in Fig.~\ref{fig:MPC testing overview}, both the robustness functionality and the SA functionality happen as secure MPCs, i.e. using cryptographic techniques that hide the computation results and all intermediary values. 
In \cite{prevwork}, we demonstrated how to compute robustness in MPC and evaluated it empirically. In this paper, we do the same for solving \eqref{eq:falsification} with SA.

\textbf{Simulated annealing (SA).}
SA is iterative: it produces a sequence of pairs $(\vec{c}_1, \inpSig_1), (\vec{c}_2, \inpSig_2),\ldots$, with $(\vec{c}_i,\inpSig_i) \in \inpSet^m\times \inpSet^m$, s.t. inputs $\inpSig_1, \inpSig_2, \dots$ converge almost surely to the global minimizer of $\rho(\cdot,\formula)$ \cite{kirkpatrick1983optimization}. In practice, SA is interrupted as soon as a falsifier is found, i.e. $\inpSig$ with negative $\rho(S(\inpSig),\formula)$.
The initial $\inpSig_1$ is drawn from some distribution over $\inpSet^m$. 
The initial \emph{candidate} $\vec{c}_1$ is produced by running steps 4-6 of the following with $k=0$.
Assume $(\vec{c}_1, \inpSig_1),\ldots,(\vec{c}_k, \inpSig_k)$ have been generated, and let $r_k :=\rho(S(\inpSig_k),\formula)$.  To generate $\vec{c}_{k+1}$ and $\inpSig_{k+1}$, SA
\begin{enumerate}
    \item \label{s4} computes robustness $\gamma_{k+1}:=\rho(S(c_{k}),\formula)$
    \item \label{s5} samples a random scalar positive \textit{temperature} $\tau_{k+1}$. 
    \item \label{s6} The candidate $\vec{c}_k$ is accepted with probability proportional to $\min\{1,e^{(r_k-\gamma_{k+1})/\tau_{k+1}}\}$. If accepted set $\inpSig_{k+1} := \vec{c}_{k}$, else $\inpSig_{k+1} := \inpSig_{k}$.
    \item \label{s1} samples a random unit-length direction $\vec{d}$ in $\reals^m$
    \item \label{s2} computes the longest line segment $L$ passing through $\inpSig_k$, contained in $\inpSet^m$, along direction $\vec{d}$
    \item \label{s3} samples $\vec{c}_{k+1}$ uniformly at random from $L$
    \item Set $k \leftarrow k+1$ and goto 1.
\end{enumerate}
Steps 1-6 constitute one iteration.
Executing all iterations in one MPC instance would require a prohibitively large circuit and large runtime. (To execute our functionality in MPC, the first step is to compile it into an arithmetic circuit, which involves unrolling all the iterations). 
So instead, we choose to set each iteration onto a fresh MPC instance: the circuit for one iteration is much smaller, and compiles and executes faster.
The inputs to the instance for iteration $k$ are $\inpSig_k, r_{k}, \vec{c}_k$ and $\rho(S(\inpSig_k),\formula)$.

\begin{table}[t]
\caption{Privacy of data.\label{tab:privacy}}
\centering
\begin{tabular}{|c|c|c|c|}
\hline
Public & \makecell{Designer Only} & \makecell{Tester Only} & Hidden \\
\hline
SA temperature & SUT input & formula & $\rho(S(\vec{c}_k),\formula)$ \\
SUT input bounds & SUT output & & anchor $\inpSig_k$ \\
& & & $\rho(S(\inpSig_k),\formula)$ \\
\hline
\end{tabular}
\end{table}

\textbf{Loop data.}
There are eight pieces of data passed through the testing loop. They are all referred to in Table \ref{tab:privacy}.
The first column of the table contains public data known by both parties: SA temperature schedule $(\tau_k)$ and SUT input bounds $\inpSet^m$. 
The temperature is independent of any party's data (in the probability sense) and chosen in advance of the testing loop. 
The bounds $\inpSet^m$ are also set in advance of the testing loop. They are known to Designer since they are derived from the design's characteristics.
Having Tester know them does not significantly increase Tester's knowledge of the SUT, considering that they do not learn the actual inputs. 

The second and third column contain Designer's and Tester's private data respectively. 
The last column shows the data that is hidden from both parties: the candidate point's robustness $\rho(S(\vec{c}_k),\formula)$, the anchor point $\inpSig_k$, and the anchor point's robustness $\rho(S(\inpSig_k),\formula)$. These three pieces of data are derived (indirectly) from private data, and thus must stay hidden from both parties.

Recall that each SA iteration gets its own fresh MPC instance. 
SA requires that three pieces of data are preserved between iterations -- the candidate point $\vec{c}_k$, the anchor point, and the anchor point's robustness.
(The SUT input is known by the Designer so they can provide it as input in the new iteration).
To do this,
we can use \emph{secret sharing} \cite{shamir1979share}, a cryptographic technique that allows each party to hold onto parts of the data without learning what it is. Secret sharing is provided by MP-SPDZ, allowing for data to persist between instantiations of MPC black boxes.
By hiding the data between iterations, we can select or reject the candidate point without revealing anything about the robustness values -- not even their ordering -- a challenge of using MPC for SA.

We can also use secret sharing as a way to pass data between the robustness MPC box and the SA MPC box -- MP-SPDZ allows us to output the computed robustness value directly as a secret share, and take it as input for the SA, hiding it from both parties in transit.
Thus in total we store three pieces of data in secret sharing: the anchor, anchor's robustness, and candidate's robustness (computed during the iteration).

\textbf{Improving efficiency.}
We can further decrease runtime by letting Tester compute the random direction $\vec{d}$ from Step \ref{s1}: this is just a vector in the unit sphere, and reveals nothing about the private data. Tester computes this in the clear and provides it as input to the MPC, which is faster than computing it cryptographically in MPC. Because our security model assumes semi-honest parties that follow the protocol (see Definition \ref{def:intuitive security}), it is assumed that Tester will not maliciously manipulate $\vec{d}$ to learn additional things it shouldn't.

\begin{figure}
    \centering
    \includegraphics[width=0.7\linewidth]{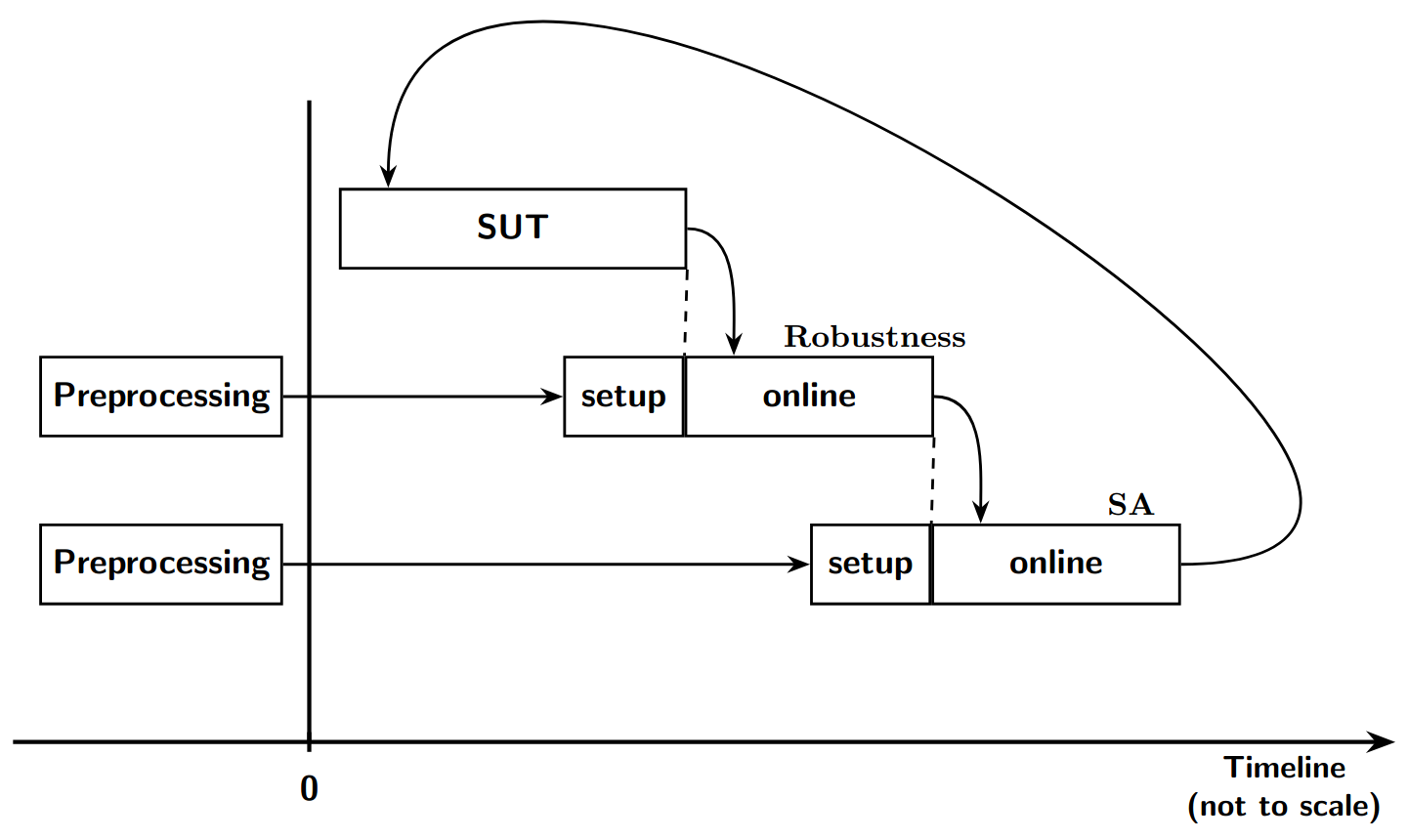}
    \caption{Concurrent phases. Preprocessing can occur before testing begins, whereas Setup can occur during the SUT simulation. Required sequential operations are first the SUT simulation, then online MPC robustness calculation, followed by online MPC SA.}
    \label{fig:loop timeline}
\end{figure}

\textbf{Concurrent phases.}
In MP-SPDZ, an MPC black box has three phases: Preprocessing, Setup, and Online.
By running certain phases concurrently, it is possible to further reduce runtime.
The Online phase is the execution of the functionality (e.g. robustness calculation or SA); this is the only phase that requires the private data (formula and trace).
The Preprocessing phase generates random data for the MPC computation, independently of the parties' inputs. 
MP-SPDZ allows for this preprocessing data to be computed and stored on disk to be used later, so we can perform this phase well in advance of the testing loop.
The Setup phase is a preparation for running the Online phase and cannot be pre-executed, but it can be run concurrently with Online phase of another functonality. 
In Fig.~\ref{fig:loop timeline} we show how
the robustness Setup can be run while the SUT is running, and the SA Setup can be run while the robustness Online phase is running.

\textbf{Communication cost} is dependent on the MPC protocol used. We use a semi-honest version of the MASCOT protocol, so we refer the reader to \cite{cryptoeprint:2016/505} for details. 
The cost does not depend on the particular inputs used, only their sizes, otherwise the cost could itself leak something about the secret inputs, so the numbers in \cite{cryptoeprint:2016/505} are applicable to our setting.

\section{Experimental Results}
We aim to show that simulated annealing in MPC is not impractically slower than simulated annealing \emph{in the clear} (i.e. not private). We compare runtimes for a single iteration of SA.
The in-the-clear implementation is written in Python and computes temperature and random direction on-the-fly, as done in standard implemenations, because their runtimes are insignificant portions of the work done in the SA computation.
The private implementation uses the MP-SPDZ compiler via the Python interface, and runs with the semi-honest MPC protocol MASCOT \cite{cryptoeprint:2016/505}. The private implementation pre-computes the temperature schedule $(\tau_k)$ and random directions and feeds them as inputs to the Online phase.

Experiments were run on a machine with a AMD Ryzen 7 PRO 7840U CPU and 16 GB of LPDDR5 RAM using MP-SPDZ version 0.4.2. The SUT was a bicycle model, with two types of controls: steering and acceleration. The output trace was of length 10, representing the bicycle's position along one axis every 5 seconds for 50 seconds total. The formula was $G_{[0, 20]} (x \leq 20 \lor x > 25) \lor F_{[21,30]} (x \leq 0 \lor x > 7)$ -- \emph{``Either position $x$ is at most 20 or greater than 25 throughout the first 20 seconds, or sometime between 21 and 30 seconds it is non-positive or greater than 7''}. 
The SA temperature schedule was set to a geometric series $\tau_k = 100\times 0.98^k$. MPC fixed-point operations were 37 bits.
\textit{The specific formula, SUT, and output trace length do not affect SA runtimes}; the only inputs to SA are robustness and a vector of control points $\inpSig$, and the only output is another vector of control points. Therefore we don't vary these three experimental factors.

\begin{figure}
    \centering
    \includegraphics[width=0.9\linewidth]{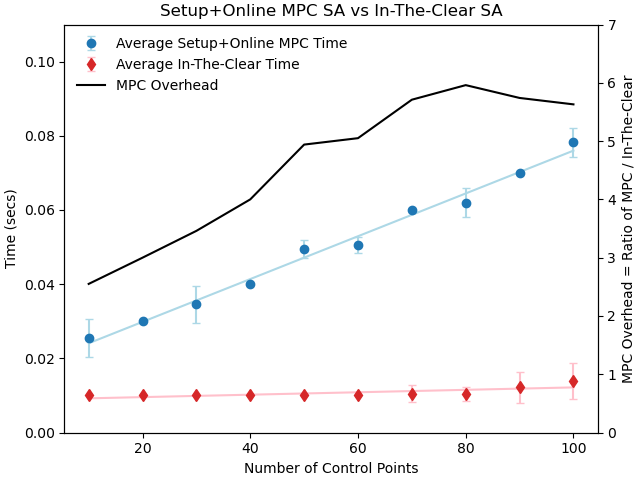}
    \caption{Experimental results comparing setup+online MPC SA with in-the-clear SA. MPC times are greater, and grow at a greater rate, than in-the-clear times as the number of control points increases, but the MPC overhead (the ratio MPC / in-the-clear) appears to saturate at the factor of 6x. Both implementation runtimes appear to grow linearly as control points increase. Points are averages of 100 SA iterations of SA at 0.01 second resolution.}
    \label{fig:runtime results}
\end{figure}

Runtimes are shown in Fig.~\ref{fig:runtime results}.
Each data point in the figure is an average over 100 iterations. Standard deviation error bars indicate low variance in these results.
These implementations are not optimized for best timing results; rather, they were chosen to be straightforward implementations of the computations. 
Further optimizations are available -- for example, MP-SPDZ has a C++ interface.
Even considering our naive implementation, the results in Fig.~\ref{fig:runtime results} suggest that private testing is practical: the key metric is the MPC overhead relative to how long it takes to simulate the system with one input (Fig.~\ref{fig:loop timeline}).
The MPC implementation, in these experiments, takes tens of milliseconds, and is at worst 6 times slower than the in-the-clear implementation. This is noticeable but is unlikely to factor significantly into the overall time of testing. 
Most SUTs take longer than tens of milliseconds to run. E.g. a PSPICE circuit or a Hardware-in-the-Loop simulation easily takes minutes to hours. 
In addition, the time taken for MPC increases as the number of control points (size of the SUT input) increases, but it appears to have a linear scaling; thus even with significantly more control points, the MPC implementation would be expected to be relatively efficient. 
Future work will perform a comprehensive set of experiments.


\bibliographystyle{IEEEtran}
\bibliography{IEEEabrv,references}

\end{document}